# "Digitwashing": The Gap between Words and Deeds in Digital Transformation and Stock Price Crash Risk

Shutter Zor
Accounting Department, Xiamen University
Xiamen, Fujian, P. R. China
Shutter_Z@outlook.com

**Abstract:** The contrast between companies' "fleshy" promises and the "skeletal" performance in digital transformation may lead to a higher risk of stock price crash. This paper selects a sample of Shanghai and Shenzhen A-share listed companies from 2010 to 2021, empirically analyses the specific impact of the gap between words and deeds in digital transformation (*GDT*) on the stock price crash risk, and explores the possible causes of *GDT*. We found that *GDT* significantly increases the stock price crash risk, and this finding is still valid after a series of robustness tests. In a further study, a deeper examination of the causes of *GDT* reveals that firms' perceptions of economic policy uncertainty significantly increase *GDT*, and the effect is more pronounced in the sample of loss-making firms. At the same time, the results of the heterogeneity test suggest that investors are more tolerant of state-owned enterprises when they are in the *GDT* situation. Taken together, we provide a concrete bridge between the two measures of digital transformation - digital text frequency and digital technology share - and offer new insights to enhance capital market stability.

## 1. Introduction

Digital transformation is a process that aims to improve an entity by triggering significant changes to its properties through combinations of information, computing, communication, and connectivity technologies (Bhimani and Willcocks, 2014; Vial, 2019). Also, it is a concrete measure aimed at achieving high-quality economic development, as highlighted by the Chinese government in a number of long-term plans.

It is thus clear that digital transformation occupies an important strategic position in the process of digital reform in micro-enterprises and in the process of high-quality development of the macro-economy. Based on this reality, some scholars have studied the economic consequences of digital transformation and have argued that digital transformation can affect the capital market performance of companies (Jiang et al., 2022; Wu et al., 2022).

In terms of measuring digital transformation, existing studies can be broadly grouped into three categories: (1) The proportion of terms related to digital transformation in the annual report (Ma et al., 2022; Tu and He, 2023; Zhu et al., 2023), which refers to the frequency of words that reflect digital transformation in the MD&A (Management Discussion and Analysis) section of annual reports, usually scaled by the length of MD&A part. (2) The share of digital assets, usually measured by the proportion of digital technology in intangible assets or total assets disclosed in the notes to the annual report (Ravichandran et al., 2017). (3) The existence of digital transformation actions, usually measured by binary variables (Gaglio et al., 2022). Overall, these methods are based on the distillation of the information content of the text and the summation of the data. Inevitably, there are differences in the calculation of the specific amount of digital transformation between different methodologies and, at the same time, these differences are rarely discussed in the relevant literature. This discrepancy, we believe, may reflect the difference between what companies say and do in terms of digital transformation.

This difference in metrics caused by the measurement methodology is an indicator of the gap between words and deeds in digital transformation. On the one hand, in terms of the "words" of digital transformation, the level of digital transformation is measured by the words related to digitalization in texts such as annual reports, which reflect more the willingness of companies to transform digitally. On the other hand, in terms of the "deeds" of digital transformation, the level of digital transformation is measured by the actual technology investment of the company, which is more reflective of the concrete behavior of the digital transformation. Based on these two separate perspectives, we developed the concept of *GDT* (the **g**ap between words and deeds in **d**igital **t**ransformation).

According to relevant literature, some researchers argued that digital transformation can affect the capital market performance of companies, i.e., it can affect the stock price crash risk (*SPCR*, thereafter) (Jiang et al., 2022; Song, 2022; Wu et al., 2022). These scholars suggested that proper digital transformation can reduce the *SPCR* of a company. The prerequisite for this is that their measures of digital transformation can be an accurate reflection of the extent of digital transformation in companies. However, based on the previous analysis, it can be found that there is a *GDT* phenomenon in the business. From both *GDT* perspectives, the "words" and "deeds" of digital transformation can individually have a positive impact on the business, but when combined with *GDT*, the inconsistency between words and actions does not necessarily have a positive impact on the business. In the light of the available research, it is not possible to know from the existing literature the impact of corporate *GDT* on the *SPCR*. Therefore, this paper attempts to provide support for the relationship between *GDT* and

*SPCR* and aims to enrich the literature in the related field.

This study makes three important contributions. First, we provide evidence that digital transformation can reduce a firm's *SPCR*, but that this effect will be very limited when there is inconsistency between what firms say and do, and even can significantly raise the *SPCR*. This is an aspect that is often overlooked by researchers and is one of the possible shortcomings of the textual analysis approach in accounting field. Our research extends the previous literature and contributes to the existing research articles on digital transformation. Second, we propose a new way that can measure the extent to which a company's digital transformation does not match its words with its actions. In contrast to previous research, this paper does not look at the "words" or "deeds" of digital transformation in isolation, but rather combines the two approaches to calculate firms' *GDT*. Third, we contribute to the literature on corporate behavior and inconsistent theory (Du, 2015a; Du, 2015b). Our study explains the impact of *GDT* on *SPCR* while also exploring the causes of *GDT*, which will help other scholars to gain a clearer understanding of the motivations behind particular corporate behaviors.

## 2. Theoretical background and hypothesis derivation

*SPCR* is closely related to the capital market stability, and while there are a number of determinants of the triggering factors, in general, the factors that influence the *SPCR* can be grouped into the following five categories (Habib et al., 2018): (1) financial reporting and corporate disclosure, (2) managerial incentives and managerial characteristics, (3) capital market transactions, (4) corporate governance mechanisms, and (5) informal institutional mechanisms. Existing studies mostly generalize these five categories of determinants in terms of agency theory (Ben-Nasr and Ghouma, 2018), information transmission theory (Kim et al., 2019), information asymmetry theory (Callen and Fang, 2015; Kim and Zhang, 2016), etc., and point out that managers' self-interest motivation and disagreement with firm owners as aligned interests are benchmark elements that trigger *SPCR*. These theories suggest that the separation of business and ownership leads to a potential conflict of interest between the managers and the owners of the company, which gives the managers an incentive to hide the true situation of the company. At the same time, the "pay-for-performance" and "pay-for-share" linkages give operators an incentive to promote a better corporate image to investors, i.e., to exaggerate the company's multiple performance and actual investment in return for more investment and more profit. In the short term, information disseminated by managers through various channels, such as annual reports, that does not reflect the true situation of the company may attract a certain amount of investment and make managers profitable, but the bubble hidden in the information also increases the *SPCR*.

In the era of Industry 4.0, the digital economy is booming, bringing with it many opportunities and challenges (Lee et al., 2014; Khan et al., 2017; Li et al., 2019). More and more companies are looking forward to riding the wave of digital transformation through the application of digital technologies, and are increasingly active in promoting their digital transformation strategies (Frank et al., 2019; Newman et al., 2021; De Bem Machado et al., 2022). The implementation of a digital transformation strategy can enhance a company's mastery of its own data and information indicators, thereby improving its performance (Zhai et al., 2022), enhancing the comparability of accounting information (Bertomeu and Marinovic, 2016; Nie et al., 2022; Xu et al., 2022), improving its capital market performance (Jabłoński, 2018; Liu and Liu, 2023), and reducing the *SPCR* (Jiang et al., 2022; Wu et al., 2022). However, it is worth noting that most of this literature uses a textual word frequency approach to construct digital transformation metrics for firms, and subject to the limitations and plausibility of the measurement methodology, it is difficult to define whether it is the actual digital transformation actions of firms that bring about these positive benefits or the spillover effects of the external rhetoric of firms claiming digital transformation, i.e., firms may benefit through the invisible promise of a digital transformation strategy. At the same time, there are a number of individual investors in the Chinese capital market who have an aggregation effect and tend to exhibit herding behavior (Tan et al., 2008; Yao et al., 2014; Xu et al., 2023). On such a realistic basis, they will over-represent the relevant information in annual reports, leading to an increase in the *SPCR*.

In conclusion, when the above theoretical background is combined with the findings of existing research, it can be seen that our analysis contradicts existing research, which is the subject of this paper. In our opinion, this discrepancy is due, on the one hand, to a lack of knowledge about the indicators of digital transformation and the inappropriate use of textual analysis methods in accounting field. On the other hand, it is the possible existence of the *GDT* phenomenon in companies. When the *GDT* phenomenon is present in a company, it can manifest itself specifically in the form of companies overstating the extent of their digital transformation, which involves misrepresenting the digital transformation. This misrepresentation leads investors to make the wrong decisions in the short term, but behind the high returns lie high risks, and a bubble in the company's share price builds up. Therefore, we believe that the *GDT* phenomenon will lead to an increase in companies' *SPCR*. Furthermore, the only hypothesis of this paper is based on this.

Hypothesis: All other things being equal, the presence of the *GDT* phenomenon in companies will have a positive effect on their *SPCR*.

## 3. Data and Methodology

Data used include: (1) MD&A texts of annual reports of Chinese A-share listed companies from Chinese Research Data Services (CNRDS), (2) other financial/non-financial indicators from China Stock Market & Accounting Research (CSMAR), and (3) collected manually. For the sample companies, and in line with the available literature, we selected Chinese companies listed on the A-share market for the period 2010 to 2021 and performed some processing as following. First, we excluded the sample of financial companies with industry codes starting with "J". Second, we excluded non-regularly quoted companies, ST, *ST and companies delisted during the period. Third, we excluded companies that went public between 2010 and 2021. Fourth, we performed a two-sided tailing of 1% on each side for all continuous variables before entering the calculation and regression.

*3.1. Definition of key variables*

3.1.1. Stock price crash risk (*SPCR*)

Drawing on Bauer et al. (2021) and Bao et al. (2022), this paper uses the negative coefficient of skewness (*NCSKEW*) to measure the stock price crash risk. We follow extensive prior research and calculate the *NCSKEW* based on each company's weekly individual stock returns as well as the corresponding weekly current price weighted market returns. Simultaneously, in order to improve the robustness of the measure, we draw on some literature in the field of equity investment and introduce market rates of return with different weightings (Pae and Sabbaghi, 2015; Qin and Singal, 2022). Finally, we have calculated the *SPCR* on the basis of two other market returns (equally weighted and total price weighted), called *SPCR1* and *SPCR2*. And, these various indicators will be discussed in the section entitled "Robustness tests".

3.1.2. The gap between words and deeds in digital transformation (*GDT*)

The *GDT* is a measure of the "difference" between the digital transformation text mentioned in the MD&A section of a company's annual report and the digital transformation actions actually taken. It shows whether companies are "talking" more about digital transformation or "acting" more, and whether they are misrepresenting the implementation of their digital transformation strategy.

In existing research, scholars have mostly defined the extent of digital transformation in terms of a single dimension of "words" and "deeds", but few scholars have considered the difference between the extent of digital transformation reflected in companies' annual reports and the actual extent of digital transformation. Methodologically, the two approaches of "words" and "deeds" explain the extent of digital transformation from different perspectives, but it is difficult to define whether companies are suspected of exaggerating the extent of digital transformation. To validate this digital transformation hype, i.e., the digital transformation gap between "words" and "deeds", this paper proposes the concept and measurement of *GDT*, drawing on the approach of earnings management in accounting field (Schipper, 1989; Chung et al., 2002; Xie et al., 2003; Myers et al., 2007; Karjalainen et al., 2023).

**Measuring digital transformation in "words" perspective (*DTW*).** Literally, the digital transformation "word" is the content of the MD&A section of the annual report that contains words related to digital transformation. Based on existing practice (Li et al., 2022; Tian et al., 2022; Zhu et al., 2022; Tu and He, 2023;), we selected the Digital Transformation Dictionary (76 words) to filter the MD&A section of the annual report by matching regular expressions. In addition, taking into account possible differences in the length (the total number of words) of the text in the MD&A section would make comparisons between companies difficult. We therefore scaled the total number of digital transformation words by the length of the MD&A section.

**Measuring digital transformation in "deeds" perspective (*DTD*).** Similarly, the digital transformation "deeds" is the content of the total invested resources that contains resources related to digital transformation. Investing in digital transformation can unlock unique resources for organizations. And the resource-based view assumes that a firm's unique resources give it a competitive advantage (Wernerfelt, 1984; Kraaijenbrink, 2010; Okorie, 2023). In this way, companies will basically tend to implement some real digital transformation measures. As digital transformation is mostly software-based, we use the percentage of digital assets in intangible assets to measure the true digital transformation investment, i.e., the digital transformation "deeds".

**The gap between words and deeds in digital transformation (*GDT*).** Borrowing ideas from earnings management (Peasnell et al., 2000; Cao et al., 2023; Cheng et al., 2023; Dimmock et al.,2023; Rahman et al., 2023), we use regression to predict the potential number of possible *DTDs* for next year based on the current year's *DTW*, called *DTDhat*. We then subtract the *DTD* from the *DTDhat* to get the *GDT*, which represents the portion of the business that should have been invested in digital transformation in the following year, but was not (also, the gap). In this regression, we also controlled for sentiment variables in the MD&A, total words, and individual, year fixed effects.

Figure 1 shows the distribution of mean data for the above three indicators by year of observation sample. It is also possible to observe the change in *GDT* calculated every four years using the images in the appendix.

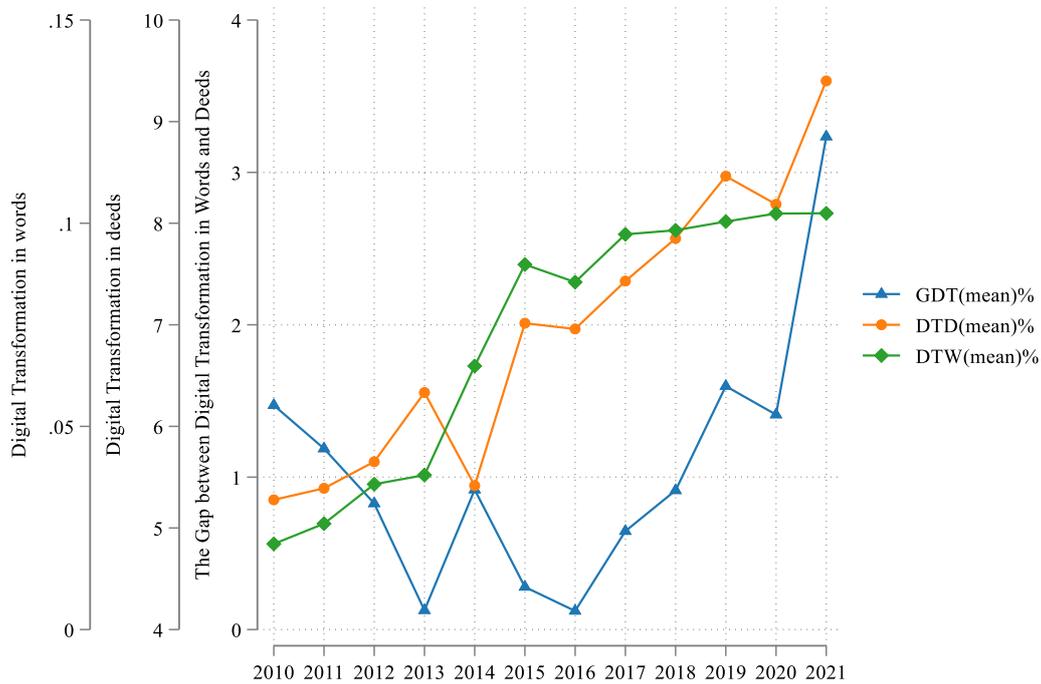

Figure 1. Development of the three indicators.

3.1.3. Control variables

Based on the available literature, we selected a reasonable set of control variables to allow our model to better reflect the statistical relationship between *GDT* and *SPCR*. The definitions and symbols of the control variables are given in Table 1.

Table 1. Definition of variables.

| Variable | Definition |
|---|---|
| BM | Ratio of book value to market value |
| Age | Natural logarithm of the age of the enterprise from the year of birth |
| Lev | Ratio of total liabilities to total assets |
| ROA | Ratio of net profit to total assets |
| Size | Natural logarithm of total assets |
| Growth | Annual growth rate of operating income |
| TobinQ | Tobin's Q, equals the market value of a company divided by its assets' replacement cost. |
| Cashflow | Ratio of cash and cash equivalents to total assets |
| Audit | Dummy variable = 1 if the company has received an unqualified audit opinion |
| Big4 | Dummy variable = 1 if the company has been audited by a Big 4 accounting firm |
| Dual | Dummy variable = 1 if the chairman and the managing director of the company are the same person |

In addition to these control variables in Table 1, we also include individual fixed effects, year fixed effects and industry fixed effects in subsequent regression models to mitigate endogeneity issues.

*3.2. Regression specifications*

To analyze whether the *GDT* affects a company's *SPCR*, we estimate the following regression results in a general form on a long panel for the period 2010-2021:

$$SPCR_{i,t} = \alpha + \beta GDT_{i,t} + \gamma X_{i,t} + \delta_i + \varphi_t + \lambda_{ind} + \varepsilon_{i,t} \qquad (1)$$

The dependent variable $SPCR_{i,t}$ represents the stock price crash risk of firm *i* in year *t*. The independent variable $GDT_{i,t}$ represents the gap of words and deeds in digital transformation of firm *i* in year *t*. In our specified regression model, *β* is the coefficient we need to focus on, which implies the size of the statistical effect of *GDT* on *SPCR*. In addition, the variable $X_{i,t}$ is a vector of control variables. Following existing research (Bedendo et al., 2023; Dutordoir et al., 2023), our specifications include firm fixed effects, $\delta_i$, which control for all firm-level time-invariant factors that might affect the *SPCR*. We also include year fixed effects, $\varphi_t$, which control for any year-level firm-invariant factors that might affect the *SPCR*. Finally, we cluster standard errors at the industry level ($\lambda_{ind}$).

## 4. Empirical results and analysis

*4.1. summary statistics*

Table 2 shows basic summary statistics for our sample of enterprise panels for the period 2010-2021. Our key explanatory variable is the *GDT*, which reflects the gap between what companies are saying and what they are doing in terms of digital transformation. In the summary statistics in Table 2, a *GDT* greater than zero indicates that companies say more about digital transformation than they do. This means that companies are saying a lot about digital transformation, but not actually doing as much. At the same time, the median of GDT is 0.048, meaning that at least 50% of the observed sample has more words than deeds when it comes to digital transformation. The standard deviation and mean of *GDT* are 0.176 and -0.003 respectively, which also indicates that they have a large variation in *GDT* among all our observed samples.

Table 2. Summary statistics.

| Variables | N | Mean | Std. Dev. | Min | P25 | Median | P75 | Max |
|---|---|---|---|---|---|---|---|---|
| *SPCR* | 10344 | 0.036 | 0.805 | -2.148 | -0.486 | 0.066 | 0.504 | 1.389 |
| *GDT* | 10344 | -0.003 | 0.176 | -0.929 | 0.000 | 0.048 | 0.065 | 0.110 |
| *BM* | 10338 | 1.518 | 1.486 | 0.128 | 0.574 | 1.030 | 1.891 | 8.268 |
| *Age* | 10338 | 3.015 | 0.251 | 2.398 | 2.833 | 3.045 | 3.219 | 3.497 |
| *Lev* | 10338 | 0.494 | 0.187 | 0.081 | 0.357 | 0.505 | 0.637 | 0.863 |
| *ROA* | 10338 | 0.042 | 0.048 | -0.096 | 0.014 | 0.033 | 0.062 | 0.212 |
| *Size* | 10338 | 22.841 | 1.346 | 20.212 | 21.867 | 22.712 | 23.689 | 26.592 |
| *Growth* | 10338 | 0.147 | 0.358 | -0.510 | -0.021 | 0.093 | 0.231 | 2.208 |
| *TobinQ* | 10176 | 1.829 | 1.194 | 0.818 | 1.107 | 1.428 | 2.054 | 7.912 |
| *Cashflow* | 10338 | 0.051 | 0.068 | -0.147 | 0.012 | 0.049 | 0.091 | 0.242 |
| *Audit* | 10338 | 0.986 | 0.119 | 0.000 | 1.000 | 1.000 | 1.000 | 1.000 |
| *Big4* | 10338 | 0.099 | 0.298 | 0.000 | 0.000 | 0.000 | 0.000 | 1.000 |
| *Dual* | 10338 | 0.148 | 0.356 | 0.000 | 0.000 | 0.000 | 0.000 | 1.000 |

In addition, we tested the SPCR performance of two different groups, those with *GDT* greater than 0 (Group1) and those with *GDT* less than 0 (Group2), in Table 3.

These two groups represent whether words are greater than actions or words are less than actions in terms of digital transformation, respectively.

Table 3. Between-group difference test.

| Group | N | Mean of SPCR | Median of SPCR | Homogeneity of variance | Median difference |
|---|---|---|---|---|---|
| 1 | 8663 | 0.035 | 0.066 | SD(Group2)/SD(Group1) | 0.043*** |
| 2 | 1681 | 0.044 | 0.023 | P = 0.013 | |

Note: *, **, and *** indicate statistical significance at the 10%, 5%, and 1% levels, respectively.

According to the results in Table 3, among the digital transition proxies in this paper, the number of samples in which "words" is greater than "deeds" is 8663, which far exceeds the number of samples in which "deeds" is greater than "words" by 1681.This means that when a t-test is used to estimate the difference in means between these two groups, the results of the variance homogeneity between groups must be considered first (Moser and Stevens, 1992; Osborne, 2010; Huang et al., 2022). The results of the variance homogeneity test significantly rejected the use of the t-test ($p$=0.013), i.e., it was not possible to conclude from the mean that Group1 had a lower risk of a share price crash.

However, according to the results of the median difference test (Murgulov et al., 2019; Goyal and Park, 2002; Zaman et al., 2021), it is easy to see that the median value (of *SPCR*) for the sample of observations where "words" are greater than "deeds" (Group1) is significantly higher than that for the sample of observations where "deeds" are greater than "words" (Group2) in digital transformation. The median difference is 0.043 and is significant at the 1% level of significance. This can be interpreted as an indication that Group1 had more *SPCR* than Group2. Namely, companies that talk more about digital transformation than they do tend to have a higher risk of share price collapse.

*4.2. Impact of GDT on SPCR*

Table 4 reports the results of the baseline regression of the relationship between *GDT* and *SPCR*. In column (1), when we include only individual fixed effects, time fixed effects and industry clustering effects, the coefficient of the effect of *GDT* on *SPCR* is 0.075 (*coefficient*=0.075, *t*=3.06, *p*<0.01). In other words, for every unit increase in the *GDT*, there is approximately a 7.5% increase in the risk of a company's stock price crash. In column (2) we include a number of other control variables that may have an impact on the relationship between *GDT* and *SPCR*. In this case, the coefficient of the effect of GDT does not change, but the significance is slightly improved (*coefficient*=0.075, *t*=3.25>3.06). Furthermore, the goodness of fit of the model improved from 0.853 to 0.858, indicating an improvement in the interpretability of the model. On this basis, the hypothesis put forward in this paper is first tested, namely that the greater the degree of inconsistency between the words and actions of a company's digital transformation, the greater the risk of a stock price crash faced by the company.

Table 4. The baseline regression of *GDT* on *SPCR*.

| | (1)<br>SPCR | (2)<br>SPCR |
|---|---|---|
| GDT | 0.075*** | 0.075*** |
| | (3.06) | (3.25) |
| BM | | 0.013** |
| | | (2.48) |
| Age | | 0.034 |
| | | (0.42) |
| Lev | | 0.012 |
| | | (0.25) |
| ROA | | 0.286*** |
| | | (3.58) |
| Size | | -0.045*** |
| | | (-4.93) |
| Growth | | -0.011 |
| | | (-1.19) |
| TobinQ | | -0.009 |
| | | (-1.50) |
| Cashflow | | -0.028 |
| | | (-0.45) |
| Audit | | 0.016 |
| | | (0.73) |
| Big4 | | 0.016 |
| | | (0.58) |
| Dual | | -0.003 |
| | | (-0.28) |
| _cons | 0.036*** | 0.927*** |
| | (39.75) | (3.12) |
| Firm | Yes | Yes |
| Year | Yes | Yes |
| N | 10338 | 10176 |
| adj. $R^2$ | 0.853 | 0.858 |

Note: *, **, and *** indicate statistical significance at the 10%, 5%, and 1% levels, respectively. The *t*-values are in parentheses (clustering standard errors at the industry level).

*4.2. Robustness tests*

4.2.1. Replacement of explained variable

According to the previous paragraphs, in order to improve the robustness of the measure, we draw on some literature in the field of equity investment and introduce market rates of return with different weightings (Pae and Sabbaghi, 2015; Qin and Singal, 2022). We have calculated the *SPCR* on the basis of two other market returns (equally weighted and total price weighted), named *SPCR1* and *SPCR2*. The results in column (1) and column (2) of Table 5 show the regression results of *GDT* on the other two *SPCRs*, respectively. Their regression coefficients are 0.032 (*coefficient*=0.032, *t*=1.87, *p<0.1*) and 0.076 (*coefficient*=0.076, *t*=3.08, *p<0.01*) respectively, both of which can support the hypothesis of this paper. It also demonstrates the robustness of our findings.

Table 5. Robustness test: replacement of explained variable.

|  | (1) SPCR1 | (2) SPCR2 |
|---|---|---|
| *GDT* | 0.032* | 0.076*** |
|  | (1.87) | (3.08) |
| *BM* | 0.001 | 0.014** |
|  | (0.50) | (2.56) |
| *Age* | -0.047** | 0.056 |
|  | (-2.23) | (0.65) |
| *Lev* | 0.030 | 0.012 |
|  | (1.21) | (0.24) |
| *ROA* | 0.105* | 0.298*** |
|  | (1.72) | (3.59) |
| *Size* | -0.004 | -0.047*** |
|  | (-0.95) | (-4.98) |
| *Growth* | -0.012** | -0.012 |
|  | (-2.06) | (-1.23) |
| *TobinQ* | -0.008* | -0.008 |
|  | (-1.91) | (-1.30) |
| *Cashflow* | -0.021 | -0.033 |
|  | (-0.52) | (-0.52) |
| *Audit* | -0.008 | 0.011 |
|  | (-0.56) | (0.49) |
| *Big4* | 0.016 | 0.015 |
|  | (0.91) | (0.55) |
| *Dual* | -0.005 | -0.001 |
|  | (-0.77) | (-0.13) |
| *_cons* | 0.341*** | 0.921*** |
|  | (3.69) | (2.94) |

| | | | | | | |
|---|---|---|---|---|---|---|
| Firm | | Yes | | | Yes | |
| Year | | Yes | | | Yes | |
| N | | 10176 | | | 10176 | |
| adj. $R^2$ | | 0.922 | | | 0.852 | |

Note: *, **, and *** indicate statistical significance at the 10%, 5%, and 1% levels, respectively. The *t*-values are in parentheses (clustering standard errors at the industry level).

4.2.2. Replacement of explanatory variable

In the previous section explaining the extent to which digital transformation is inconsistent with words and actions, we borrowed from earnings management to calculate *GDT*. To make the explanation more robust, we replaced the original method of subtracting actual values from forecast values with *DTW* minus *DTD*, named *GDT1*. We also try to convert the *GDT* into a dummy variable that takes 1 if the "words" of the digital transformation are greater than the "deeds" of the digital transformation, and 0 if not (*GDT2*). Meanwhile, we have included the newly calculated *GDT1* and *GDT2* in the *SPCR* calculated for the three different market returns for regression, and the results are shown in Table 6. The regression results in Table 6 remain robust, and we can still conclude that "inconsistencies in firms' digital transformation (*GDT*) may contribute to the risk of a stock price crash (*SPCR*)".

Table 6. Robustness test: replacement of explanatory variable.

| | (1) SPCR | (2) SPCR1 | (3) SPCR2 | (4) SPCR | (5) SPCR1 | (6) SPCR2 |
|---|---|---|---|---|---|---|
| GDT1 | 0.053*** | 0.018* | 0.053*** | | | |
| | (3.51) | (1.73) | (3.43) | | | |
| GDT2 | | | | 0.041*** | 0.011*** | 0.039*** |
| | | | | (4.12) | (2.82) | (3.90) |
| BM | 0.014** | 0.002 | 0.014** | 0.013** | 0.001 | 0.014** |
| | (2.53) | (0.55) | (2.60) | (2.50) | (0.49) | (2.58) |
| Age | 0.037 | -0.046** | 0.058 | 0.036 | -0.047** | 0.057 |
| | (0.45) | (-2.20) | (0.68) | (0.43) | (-2.25) | (0.66) |
| Lev | 0.013 | 0.030 | 0.012 | 0.015 | 0.030 | 0.014 |

|  | (0.26) | (1.21) | (0.25) | (0.32) | (1.24) | (0.30) |
| --- | --- | --- | --- | --- | --- | --- |
| ROA | 0.287*** | 0.104* | 0.298*** | 0.293*** | 0.106* | 0.304*** |
|  | (3.64) | (1.72) | (3.64) | (3.69) | (1.75) | (3.68) |
| Size | -0.047*** | -0.005 | -0.049*** | -0.044*** | -0.004 | -0.046*** |
|  | (-5.00) | (-1.06) | (-5.06) | (-4.78) | (-0.88) | (-4.84) |
| Growth | -0.011 | -0.012** | -0.012 | -0.012 | -0.013** | -0.013 |
|  | (-1.18) | (-2.06) | (-1.23) | (-1.28) | (-2.10) | (-1.32) |
| TobinQ | -0.009 | -0.008* | -0.008 | -0.008 | -0.008* | -0.008 |
|  | (-1.53) | (-1.91) | (-1.33) | (-1.44) | (-1.88) | (-1.24) |
| Cashflow | -0.028 | -0.021 | -0.032 | -0.031 | -0.022 | -0.035 |
|  | (-0.45) | (-0.52) | (-0.52) | (-0.48) | (-0.54) | (-0.55) |
| Audit | 0.016 | -0.008 | 0.011 | 0.012 | -0.009 | 0.008 |
|  | (0.74) | (-0.57) | (0.49) | (0.57) | (-0.65) | (0.34) |
| Big4 | 0.015 | 0.016 | 0.015 | 0.017 | 0.016 | 0.016 |
|  | (0.57) | (0.90) | (0.54) | (0.63) | (0.93) | (0.60) |
| Dual | -0.003 | -0.005 | -0.001 | -0.003 | -0.005 | -0.002 |
|  | (-0.29) | (-0.78) | (-0.14) | (-0.32) | (-0.79) | (-0.16) |
| _cons | 0.961*** | 0.353*** | 0.955*** | 0.874*** | 0.327*** | 0.870*** |
|  | (3.16) | (3.69) | (2.98) | (2.99) | (3.60) | (2.83) |
| Firm | Yes | Yes | Yes | Yes | Yes | Yes |
| Year | Yes | Yes | Yes | Yes | Yes | Yes |
| N | 10176 | 10176 | 10176 | 10176 | 10176 | 10176 |
| adj. $R^2$ | 0.858 | 0.922 | 0.852 | 0.858 | 0.922 | 0.852 |

Note: *, **, and *** indicate statistical significance at the 10%, 5%, and 1% levels, respectively. The *t*-values are in parentheses (clustering standard errors at the industry level).

## 5. Advanced research

*5.1. Impact of the EPU*

The way to determine the impact of differences in what companies say and do about digital transformation is to explore the reasons why companies exaggerate the description of their actual investment in digital transformation. And in this context, the potential impact of an economic policy uncertainty (*EPU*) shock on companies is one of the important factors that cannot be ignored. Baker et al. (2016) successfully constructed an index describing economic policy uncertainty in a macro environment based on newspaper texts. However, for individual firms with sufficient specificity, this uniform, macro-applicable economic policy uncertainty is hardly suitable for all individual firms. Therefore, in order to verify the impact of economic policy uncertainty on *GDT*, we reconstructed the firm-level *EPU*, or *FEPU* for short, based on the text of the annual report (Zor, 2023b). Specifically, *FEPU* is the proportion of text in the MD&A section of the annual report that relates to economic policy uncertainty.

At the same time, we argue that loss-making firms have greater incentives to exaggerate their digital transformation behavior in the hope of maintaining investor investment. Combining economic policy uncertainty and loss-making, this paper argues that economic policy uncertainty is an important influence on the variance between what firms say and do about digital transformation, and is also moderated by whether firms are loss-making. Finally, we specify the following regression model to verify the effect of *FEPU* on *GDT*, as well as the moderating effect of *Loss*.

$$GDT_{i,t} = \alpha + \beta FEPU_{i,t} + \gamma X_{i,t} + \delta_i + \varphi_t + \lambda_{ind} + \varepsilon_{i,t} \qquad (2)$$

$$GDT_{i,t} = \alpha + \beta FEPU_{i,t} \times Loss_{i,t} + \beta_1 FEPU_{i,t} + \beta_2 Loss_{i,t} + \gamma X_{i,t} + \delta_i + \varphi_t + \lambda_{ind} + \varepsilon_{i,t} \quad (3)$$

The various implications of the above models have been described in the previous section. In both models, the variable we are interested in is $\beta$. We expect $\beta$ in model (2) to be positive. This implies that when *FEPU* increases, firms may need to make more strategic digital commitments to ensure continued investment from investors. $\beta$ in model (3) should also be positive. This implies that when a firm's *FEPU* increases, firms that are losing money are more likely to want to secure sustainable investment by making more digital commitments.

Table 7. Advanced research: a potential way to influence the *GDT*.

|  | (1) GDT | (2) GDT | (3) GDT |
|---|---|---|---|
| *FEPU* | 0.014** | 0.013* | 0.012 |
|  | (2.00) | (1.93) | (1.60) |
| *Loss* |  |  | -0.024* |
|  |  |  | (-1.68) |
| *FEPU×Loss* |  |  | 0.020* |
|  |  |  | (1.83) |
| _cons | -0.018** | -0.037 | -0.032 |
|  | (-2.23) | (-0.19) | (-0.17) |
| Controls | No | Yes | Yes |
| Firm | Yes | Yes | Yes |
| Year | Yes | Yes | Yes |
| N | 10181 | 10021 | 10021 |
| adj. $R^2$ | 0.531 | 0.535 | 0.535 |

Note: *, **, and *** indicate statistical significance at the 10%, 5%, and 1% levels, respectively. The *t*-values are in parentheses (clustering standard errors at the industry level).

Table 7 shows the results of our estimation of model (2) and model (3). For brevity, we omit reporting the coefficients of the control variables. The results of the regressions support our hypothesis and conjecture. This means that when firms' perceptions of economic policy uncertainty increase, these firms will make more commitments to digital transformation in order to invest in stability, leading to the phenomenon of *GDT* in firms (*coefficient*=0.014, *t*=2.00, *p*<0.05; *coefficient*=0.013, *t*=1.93, *p*<0.1). In the case of loss-making companies, they have a greater incentive to exaggerate this effect. Therefore, the contribution of *FEPU* to *GDT* is more pronounced for loss-making enterprises (*coefficient*=0.02, *t*=1.83, *p*<0.1).

*5.2. Subgroup analysis*

Compared to the average private company, Chinese state-owned enterprises (SOE) tend to have more resources and easier access to more bank loans and policies due to their certain political connections (Sheng et al., 2011; Guo et al., 2014; Liu et al, 2018; Wang et al., 2022). Therefore, we believe that investors have a higher tolerant threshold for the *GDT* of SOEs. This means that if the *GDT* of SOEs increases, it will have less impact on the *SPCR* compared to private companies.

The mean (*GDT (mean)*) in Table 8 suggests that, on average, SOEs tend to do more actual work (deeds rather than words) when it comes to digital transformation, while non-SOEs' descriptions of their actual investments in digital transformation are somewhat exaggerated (words rather than deeds). The test for difference between groups has a p-value of 0.056 for 500 samples and 0.057 for 1000 samples, rejecting the initial hypothesis (no difference between groups) that there is a significant difference between groups. On balance, state-owned enterprises, because of their various competitive advantages, tend to be more confident in their ability to do what they say they will do, so even if their words outweigh their deeds for a period of time, it will not lead to a significant increase in the risk of stock price crash; whereas for non-state-owned enterprises, because of the need to make profits and attract investment, saying but not doing will more easily lead to investor dissatisfaction, which will be reflected in the stock market as an increased risk of stock price crash.

Table 8. Advanced research: subgroup analysis.

|  | (1) SOE<br>SPCR | (2) Non-SOE<br>SPCR |
|---|---|---|
| GDT | 0.038 | 0.112** |
|  | (1.03) | (2.11) |
| _cons | 0.525 | 1.623*** |
|  | (1.26) | (2.66) |
| Controls | Yes | Yes |
| Firm | Yes | Yes |
| Year | Yes | Yes |
| N | 6713 | 3450 |
| adj. $R^2$ | 0.874 | 0.830 |
| GDT (mean) | -0.005 | 0.003 |
| Between-group differences | 500/0.056 | 1000/0.057 |

Note: *, **, and *** indicate statistical significance at the 10%, 5%, and 1% levels, respectively. The *t*-values are in parentheses (clustering standard errors at the industry level).

# 6. Conclusion

Digital transformation is an inevitable and important change for companies in the digital economy, and can bring better economic benefits to companies. The contrast between rich digital transformation strategies and poor digital transformation investments increases the risk of share price collapse and affects the stability of the capital markets, which in turn affects the actual benefits of companies.

This paper selects a sample of listed companies in China's Shanghai and Shenzhen A-shares from 2010 to 2021 to empirically analyze the specific impact of *GDT* on the *SPCR*, and to explore the possible causes of inconsistent words and deeds in digital transformation. We find that *GDT* significantly increases the *SPCR*, and this finding holds after a series of robustness tests. In a further study, a deeper exploration of the causes of *GDT* reveals that firms' perceptions of economic policy uncertainty significantly increase *GDT*, and the effect is more pronounced in the sample of loss-making firms. At the same time, the results of the heterogeneity test suggest that investors are more tolerant of state-owned enterprises when they are in the *GDT* situation. Such research provides new ideas for improving capital market stability and achieving high-quality economic development.

There are in fact two main policy points that can be drawn from the findings of this paper. First, given the impact that inconsistent digital transformation can have on the stability of capital markets and thus on the interests of investors, this paper argues that relevant government authorities should develop policies to regulate the permissible range of inconsistency. The current inconsistency in digital transformation may be a promise of future digital transformation, or it may be a mere "pie in the sky" to deceive investors. By limiting the threshold of excessive disclosure by companies in digital transformation, the interests of investors can be well protected and the long-term sustainable and stable development of the capital market can be promoted. Second, relevant regulatory authorities should systematically evaluate enterprises whose words and deeds differ in terms of digital transformation, punish those whose words are greater than their deeds, and provide timely encouragement and support to those whose deeds are greater than their words. Such parallel measures of rewards and punishments will help form a group of quality enterprises that talk less and do more.

In addition, our study enriches the research literature on digital transformation and also provides a new way of thinking about value. This will help investors or relevant institutions to further assess the creditworthiness of investee when considering investment plans.

## 7. Other works

The authors' other works provide endorsement for the technical support of this paper. For example, Luo and Zor (2023a), Luo and Zor (2023b) provide this article with prior knowledge for relevant empirical analyses. Related work by Zor (2023a) and Zor (2023b) provided technical support for this working paper on text analysis.

**Appendix**

The .shp file of the map is from Alibaba Cloud's DataV.GeoAtlas.

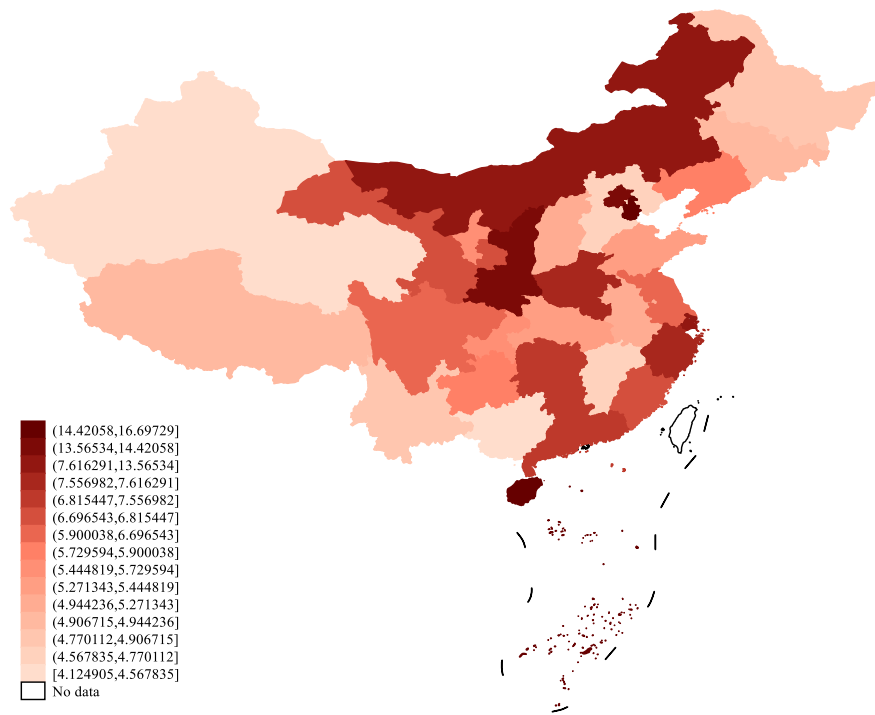

I. Geographical distribution characteristics of the average *GDT* in 2010-2013.

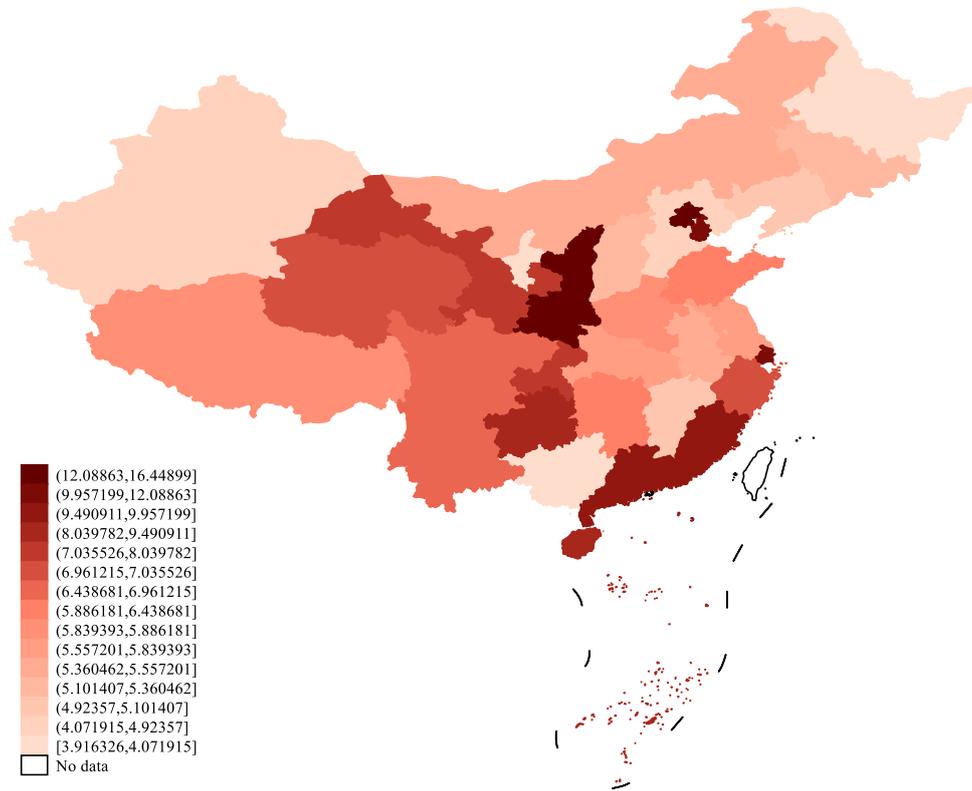

II. Geographical distribution characteristics of the average *GDT* in 2014-2017.

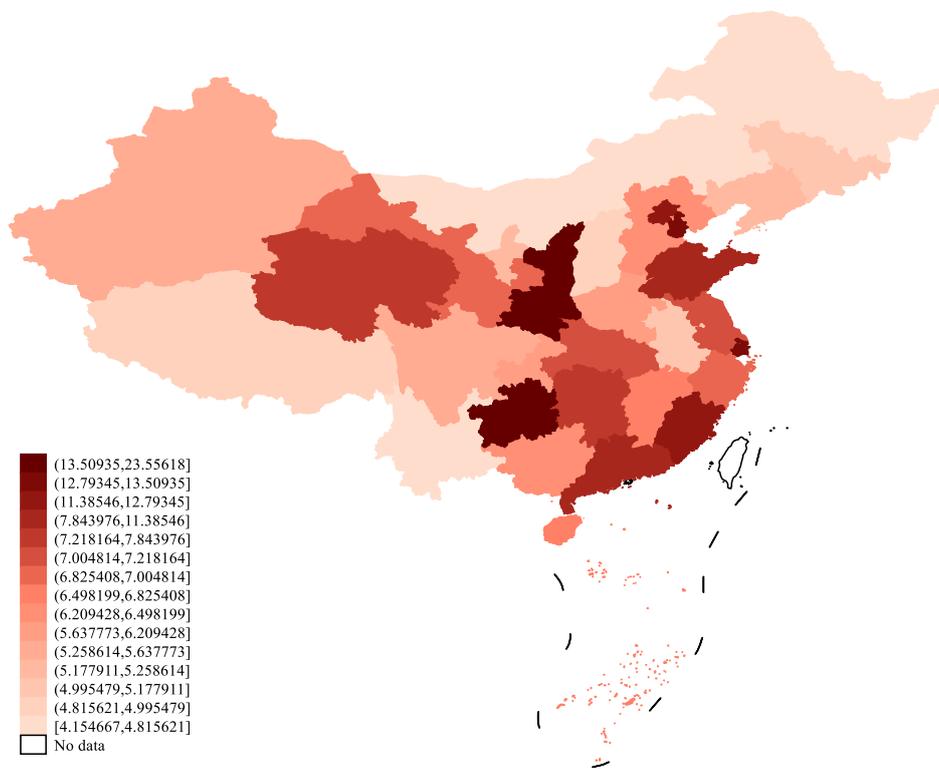

III. Geographical distribution characteristics of the average *GDT* in 2018-2021.